\documentclass{elsart}
\usepackage{graphicx,amssymb,
amstext,axodraw}
\newcounter{bla}

\begin{document}
\begin{frontmatter}

\title{FASTERD: a  Monte Carlo event generator for the study of final state radiation in the process $e^+e^-\to\pi\pi\gamma$ at DA$\Phi$NE}

\author[a,b]{O.~Shekhovtsova\thanksref{author}},
\author[a]{G.~Venanzoni},
\author[a]{G.~Pancheri}

\thanks[author]{Corresponding author}

\address[a]{INFN Laboratori Nazionale di Frascati, Frascati (RM) 00044, Italy}
\address[b]{NSC KIPT, Kharkov 61202, Ukraine}

\begin{abstract}
  %Type your abstract here.
FASTERD is a Monte Carlo event generator to study  the final state radiation
both in the
 $e^+e^-\to\pi^+\pi^-\gamma$  and $e^+e^-\to\pi^0\pi^0\gamma$ processes in the energy region of  the $\phi$-factory DA$\Phi$NE.
% \textbf{With small modification the code can be used for the energy region from the pion threshold upto the $\phi$ meson mass.}
Differential spectra that include both initial and final state
radiation and the interference between them are produced.
%Once it is done a contribution to all spectrum will be automatically done.
Three different mechanisms for the $\pi\pi\gamma$ final state are
considered: Bremsstrahlung process (both in the framework of sQED
and Resonance Perturbation Theory), the $\phi$ direct decay
($e^+e^-\to\phi\to (f_0;f_0+\sigma)\gamma\to \pi\pi\gamma$) and
the double resonance mechanism (as $e^+e^-\to\phi\to
\rho^\pm\pi^\mp\to \pi^+\pi^-\gamma$ and $e^+e^-\to\rho\to
\omega\pi^0\to \pi^0\pi^0\gamma$). Additional models can be
incorporated as well. 
\begin{flushleft}
  %Insert your suggested PACS number here
\end{flushleft}


\end{abstract}

\end{frontmatter}

{\bf PROGRAM SUMMARY}

\begin{small}
\noindent
{\em Program Title:  FASTERD}                                          \\
{\em Authors: G.~Pancheri, O.~Shekhovtsova, G.~Venanzoni}          \\
{\em Journal Reference:}                                      \\
  %Leave blank, supplied by Elsevier.
{\em Catalogue identifier:}                                   \\
  %Leave blank, supplied by Elsevier.
{\em Licensing provisions: none}                                   \\
{\em Programming language: FORTRAN77}                                   \\
{\em Computer:  any computer with FORTRAN77 compiler}                                               \\
{\em Operating system: UNIX, LINUX, MAC OSX}                                       
{\em Classification: 11.1}                                         \\
{\em External routines/libraries: MATHLIB, PACKLIB from CERN library}            \\
{\em Nature of problem: General parameterization of the
$\gamma^*\to\pi\pi\gamma$ process; test models describing
Bremsstrahlung process, the $\phi$ direct decay,
double vector resonance mechanism.            }\\
{\em Solution method: Numerical integration of analytical formulae}\\
{\em Restrictions: Only one photon emission is considered}\\
{\em Running time 28 sec with standard input card (1e6 events generated)
on a Intel Core 2 Duo 2. GHz with 1 GB RAM.}
\end{small}

\newpage

%********************************************************************
%*           INTRODUCTION
%********************************************************************

\section{Introduction}
\label{intro}

The anomalous magnetic moment of the muon ($a_\mu$) is one of the
most precise test of the Standard Model~\cite{Upgm}. Theoretical
predictions differ from the experimental result for more
than $3\sigma$~\cite{amu_th_exp}. The main source of uncertainty in the theoretical prediction comes from the hadronic contribution, $a_\mu^{(had)}$~\cite{amu_th_exp}. This contribution cannot
be reliably calculated in the framework of perturbative QCD
(pQCD), because low-energy region dominates, but it can be
estimated by dispersion relation using the experimental cross
sections of $e^+e^-$ annihilation into hadrons as an input~\cite{eid_jeg}.
About $70\%$ of the hadronic part of the muon anomalous magnetic
moment, $a_\mu^{(had)}$, comes from the energy region below
$1$ GeV and, due to the presence of the $\rho$-meson, the main
contribution  to $a_\mu^{(had)}$ is related with the $\pi^+\pi^-$
final state.

Experimentally, the energy region from threshold to the collider
beam energy is explored at the $\Phi$-factory DA$\Phi$NE,
%(Frascati, $s=4E^2=m_\phi^2$),
$PEP-II$ and $KEKB$ at $\Upsilon(4S)$-resonance
  using the method
of radiative return (for a review see~\cite{kluge} and references therein).
This method  relies
on the factorization of the radiative cross section  into the product
of the hadronic cross section times a radiation function
$H(q^2,\theta_{max},\theta_{min})$ known from Quantum
Electrodynamics (QED) \cite{Chen_75,Rr1,Rr2,Baier_65,Khoze_02}. For two pions
final state it means that, in the presence of only the initial state radiation (by leptons, ISR),
 the radiative cross section
$\sigma^{\pi\pi\gamma}$ corresponding to the process
\begin{equation}\label{process_charged}
e^+(p_+)+e^-(p_-)\to\pi^+(p_1)+\pi^-(p_2)+\gamma(k) ,
\end{equation}
can be written as  $d\sigma^{\pi\pi\gamma}=d\sigma^{\pi\pi}(q^2)H(q^2,\theta_{max},\theta_{min})$ \cite{kluge,Rr2, Khoze_02},
where $\theta_{min}$ and $\theta_{max}$ are the minimal and maximal azimuthal angles of the radiated photon, $q=p_1+p_2$ and the hadronic cross section $\sigma^{\pi\pi}$
is taken at a reduced CM energy. 
%In spite of the $\alpha/\pi$ suppression of the emission of a photon, this method has  many advantages in systematics over the more traditional direct scanning measurements performed at different CM energies, such as the experiments at VEPP--2M (Novosibirsk) \cite {cmd2} and BES (Beijing) \cite{pekin}. 
The final state (FS) radiation (FSR) is an
irreducible background in radiative return measurements of the
hadronic cross section \cite{Rr2,kloe} and spoils the factorization of
the cross section. In any experimental setup the process of FSR
cannot be excluded from the analysis. The KLOE experiment has developed two different analysis strategies:
the first one is with the photon emitted at small angle
($\theta_\gamma<15^\circ$) and the other one is for
 the photon reconstructed at large angle ($60^\circ<\theta_\gamma<120^\circ$), being for both $50^\circ<\theta_\pi<130^\circ$.
 In the case of the small
angle kinematics  the FSR contribution can be safely neglected,
while for the large angle analysis it becomes relevant (upto $40\%$ of ISR).
The large angle analysis allows to scan the pion form factor down to the threshold~\cite{kloe_large}.

Radiative corrections (RC's) related to initial
state radiation, i.e. the function $H$, can be safely computed in QED.
For the FSR process the situation is different. In the region
below $2$ GeV the pQCD is not applicable to describe FSR and
calculation of the cross section relies on the low energy
pion--photon interaction model. Thus the measured FSR cross
section gives an unique possibility to get
very interesting information on the dynamics of interacting mesons
and photons, to test the pion--photon interaction models and
extract their parameters~\cite{Pancheri:2007xt}.

In the case of neutral pions in the final states
\begin{equation}\label{process_neutral}
e^+(p_+)+e^-(p_-)\to\pi^0(p_1)+\pi^0(p_2)+\gamma(k) ,
\end{equation}
the ISR
contribution is absent\footnote{This statement is valid only if one neglects multi-photon emission},
and the cross section is determined solely
by the FSR mechanism.
This process, together with asymmetries~\cite{czyz} and cross section
in the charged channel,
allows to extract information on  pion-photon interaction
and test effective models for FSR.

For realistic experimental cuts on the angle and energy of the
final particles the cross section cannot be evaluated analytically
and one has to use Monte Carlo (MC) event generator. The first MC
describing the reaction (\ref{process_charged}) was EVA
\cite{Rr2}. EVA simulates both ISR and FSR processes for non
zero angle emitted photon ($\theta_\gamma>\theta_{min}$). For FSR
the sQED model was chosen.
Afterwards the MC PHOKHARA was written to include different
charged final state and radiative corrections to ISR
\cite{phokhara}. %As PHOKHARA was done, first, to describe the DA$\Phi$NE energy
The contribution of the $\phi$--meson direct decay, that is relevant at the DA$\Phi$NE energy, was
added, firstly in EVA~\cite{graz}  and then in PHOKHARA~\cite{czyz}.

\begin{figure}
\begin{center}
\par
\parbox{1.0\textwidth}{\hspace{-0.5cm}
%\vspace{5.cm}
\includegraphics[width=0.6\textwidth,height=0.3\textwidth]{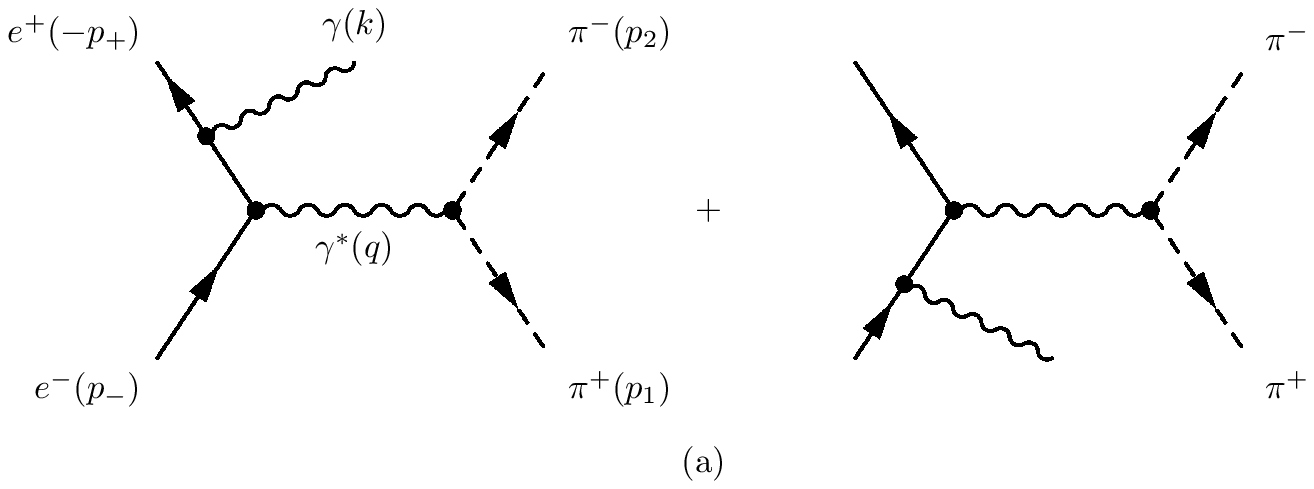}
\includegraphics[width=0.3\textwidth,height=0.25\textwidth]{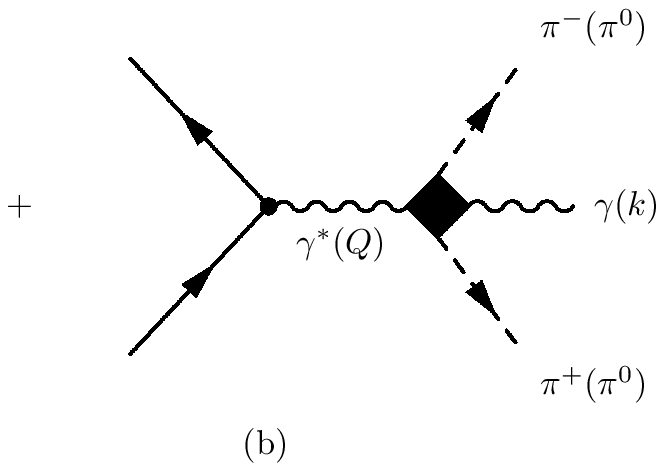}}
%\hspace{1.cm}
%%\vspace{-1.5cm}
\caption{}\label{fig_diagram}
\end{center}
\end{figure}

The  computer code FASTERD\footnote{FinAL STatE Radiation at
DA$\Phi$NE} presented in this paper is a Monte Carlo event
generator written in FORTRAN that simulates both  processes
(\ref{process_charged}) and (\ref{process_neutral}), where the
hard photon $\gamma(k)$  can be emitted by the
leptons\footnote{Only for the charged channel} and/or the
pions Fig.~\ref{fig_diagram}.
This program was inspired by EVA. The present version of FASTERD
includes different mechanisms for the $\pi\pi\gamma$
production: final Bremsstrahlung\footnote{Only for the charged channel} in the
framework of both Resonance Perturbation Theory and sQED, the
contributions related to the $\phi$ intermediate state and the
double vector resonance part (for details see
Section~\ref{fsr_model}). Up to now only the case of one photon
emission has been considered. The code contains two types of
source files: (1) the main program \textbf{fasterd.f}, where the
calculations are done,
 and (2) the input file \textbf{cards\_fasterd.dat}
which defines the parameters for the generation.
Both these files will be described in the following.
As output of the program, the cross section of the process is evaluated and
a PAW/HBOOK ntuple is produced with the 4-momenta for each event of the outgoing particles. For the convenience of the user we have added a Makefile and the test output files for the Journal library.

This paper is organized as follows. In
Section~\ref{isr} the main formulae for ISR and interference
between ISR and FSR are presented. In Section~\ref{fsr_model} we
give a general description of FSR process and present the FSR
models that are included in our program. In Section~\ref{comparison} the spectra for
different FSR models are compared with analytical results and with MC PHOKHARA. Also a possible generalization applicable to a wider region is described in Section~\ref{comparison}. In
Section~\ref{subroutine}, we summarize the general  procedure for calculating the spectrum.
In Appendices A, B, C a short description of the input and output files is presented.

%**************************************************************************
%*          INITIAL STATE RADIATION
%**************************************************************************

\section{Initial state radiation models and pion form factor}\label{isr}
The cross section of the processes (\ref{process_charged}) and
(\ref{process_neutral}) can be written as
\begin{eqnarray}
d\sigma & = &\frac{1}{2s(2\pi)^5}C_{12}\int
\delta^4(Q-p_1-p_2-k)\frac{d^3p_1d^3p_2d^3k}{8E_+E_-\omega}|M|^2  \nonumber \\
& = & C_{12} N |M|^2 dq^2 d\Omega^\gamma d\Omega^{\pi^+}, \; \; \Biggl( N =\frac{\alpha^3 (s-q^2)}{64\pi^2
s^2}\sqrt{1-\frac{4m_\pi^2}{q^2}} \; \Biggr)
\end{eqnarray}
where $\alpha$ is the fine structure constant, $m_\pi$ is the pion mass, $\omega$ is the photon energy,  $Q=p_++p_-$, $s=Q^2$ and  the invariant amplitude squared,
averaged over initial lepton polarizations
and summed over the photon polarizations\footnote{%
We use $\sum_{polar.}\epsilon _{\rho }^{\ast }\epsilon _{\sigma
}=-g_{\rho \sigma }$} is
\begin{equation}
\overline{|M|^{2}}=\overline{|M_{ISR}|^{2}}+\overline{|M_{FSR}|^{2}}+2\mathrm{Re}(%
\overline{M_{ISR}M_{FSR}^{\ast }}).  \label{sq_ampl}
\end{equation}

$M^{(ISR)}$ ($M^{(FSR)}$) corresponds to the ISR(FSR) production amplitude. The factor
$C_{12}=\mathstrut\frac{1}{2}$ for $\pi^0 \pi^0$ in the final
state and $C_{12}=1$ for $\pi^+ \pi^-$.

For ISR process the invariant amplitude squared, averaged over
initial lepton polarizations and summed over the photon
polarizations, is
\begin{eqnarray}\label{mat_isr}
&&\overline{|M^{(ISR)}|^2}=-\frac{4}{q^{2}}|F_{\pi }(q^{2})|^{2}R , \\
&&R =\frac{m_{\pi }^{2}}{q^{2}}F+\frac{\chi _{1}^{2}+\chi
_{2}^{2}-\chi
_{1}(q^{2}-t_{2})-\chi _{2}(q^{2}-t_{1})}{t_{1}t_{2}} \nonumber \\ 
&&-\frac{2m_{e}^{2}\chi_1}{t_{2}^{2}}\Biggl(\frac{\chi _{1}}{q^{2}}-1\Biggr)-\frac{%
2m_{e}^{2}\chi_2}{t_{1}^{2}}\Biggl(\frac{\chi
_{2}}{q^{2}}-1\Biggr), \text{\hspace{0.3cm}} 
F=\frac{(q^{2}-t_{1})^{2}+(q^{2}-t_{2})^{2}}{%
t_{1}t_{2}} , \nonumber
\end{eqnarray}
where $m_e$ is the electron mass, $\chi _{1,2}\equiv 2p_{-,+}\cdot p_{2}$, $t_1=-2p_-k$,
$t_2=-2p_+k$.

The nonpoint-like behaviour of pions is determined by the
form-factor (FF) $F_\pi(q^2)$ that is the function of the pion
mass squared $q^2$. In the case of the neutral channel
$M^{(ISR)}=0$. Four different parameterizations for the pion FF are
considered: K\"uhn-Santamaria (KS)~\cite{kuhn_ff},
Gounaris-Sakurai (GS)~\cite{gounaris_ff}, the RPT parametrization
and an "improved" version of K\"uhn-Santamaria (see below)~\cite{kuhn_impr}.

\subsection{K\"uhn-Santamaria and Gounaris-Sakurai pion FF}
Based on the results of Ref.~\cite{kuhn_ff}, the pion FF describing the $\rho-\omega$ mixing and the first excited $\rho$
resonance ($\rho'$), can be written as
\begin{equation}\label{ff_ks_gs}
F_\pi(q^2)=\frac{B_\rho\frac{1+\alpha
B_\omega}{1+\alpha}+\beta B_{\rho'}}{1+\beta} ,
\end{equation}
where for the KS parametrization~\cite{kuhn_ff}
\begin{equation}\label{ks}
B_r^{KS}(q^2)=\frac{m_r^2}{m_r^2-q^2-i \sqrt{q^2}\Gamma_r(q^2)}
\end{equation}
and for the GS one~\cite{gounaris_ff}
\begin{equation}\label{gs}
B_r^{GS}(q^2)=\frac{m_r^2+H(0)}{m_r^2-q^2+H(q^2)-i
\sqrt{q^2}\Gamma_r(q^2)} ,
\end{equation}
with
\begin{eqnarray*}
H(q^2)&=&\frac{m_\rho^2 \Gamma_\rho}{p_\pi^3(m_\rho^2)}\Biggl[p_\pi^2(q^2)(h(q^2)-h(m_\rho^2))+(m_\rho^2-q^2)p_\pi^2(m_\rho^2)\frac{d h}{d q^2}|_{q^2=m_\rho^2}\Biggr] , \\
h(q^2)&=&\frac{2}{\pi}\frac{p_\pi(q^2)}{\sqrt{q^2}}\ln{\frac{\sqrt{q^2}+2p_\pi(q^2)}{2m_\pi}}
, \; \; \; p_\pi(q^2)=\frac{1}{2}\sqrt{q^2-m_\pi^2} .
\end{eqnarray*}
The energy dependence for the $\rho$ mesons is taken
in the form
\begin{equation}\label{gamma}
\Gamma_\rho(q^2)=\Gamma_\rho\frac{m_\rho^2}{q^2}
\Biggl(\frac{p_\pi(q^2)}{p_\pi(m_\rho^2)}\Biggr)^3\cdot
\Theta(q^2-4m_\pi^2).
\end{equation}
For the $\omega$ resonance a simple Breit-Wigner resonance form
with constant width was used for both parameterizations.

As one can see the single resonance contribution is
normalized to unity at $q^2=0$ for both parameterizations
($B_r(0)=1$) whereas the right normalization for the pion FF,
$F_\pi(0)=1$, is realized by the corresponding choice of the
parameters $\alpha$, $\beta$.

All model parameters (the mass and width of the resonances as well
as the parameter $\alpha$, $\beta$) are determined in the input
file \textbf{cards\_fasterd.dat}.   The KS pion FF parametrization
corresponds to the function \textbf{F\_pi\_ks} whereas the GS one to
 \textbf{F\_pi\_gs}.

\subsection{RPT parametrization for the pion FF}
The Resonance Perturbation Theory is based on Chiral Perturbation
Theory ($\chi$PT) with the explicit inclusion of the vector and
axial--vector mesons, $\rho_0(770)$ and $a_1(1260)$. Whereas
$\chi$PT gives correct predictions for the pion form factor at very
low energy, RPT is the appropriate framework to describe the pion
form factor at intermediate energies ($E \sim
m_\rho$)~\cite{Ecker_89}.
According
to the RPT model the pion form factor, that describes
the $\rho-\omega$ mixing, can be written  as:
\begin{equation}\label{formfact}
F_\pi(q^2)=1+\frac{F_V G_V}{f_\pi^2}\frac{q^2}{m_\rho^2}B_\rho^{KS}(q^2)
\Biggl(1-\frac{\Pi_{\rho\omega}}{3m_\omega^2}B_\omega^{KS}(q^2)\Biggr) ,
\end{equation}
where $q^2$ is the virtuality of the photon, $f_\pi=92.4$ MeV  and the parameter $\Pi_{\rho\omega}$ describes the $\rho$-$\omega$ mixing.
As before  a constant width is used for the  $\omega$--meson. We also assume  that the
parameter $\Pi_{\rho\omega}$ is a constant and is related to the branching fraction
$Br(\omega\to\pi^+\pi^-)$:
\begin{equation}\label{br_om}
Br(\omega\to\pi^+\pi^-)=\displaystyle\frac{\mathstrut
|\Pi_{\rho\omega}|^2}{\Gamma_\rho \Gamma_\omega m_\rho^2} .
\end{equation}
As was mentioned above, the value of  $F_V$ and $G_V$,  as well
as the mass of the $\rho$ and $\omega$ mesons ($m_\rho$ and
$m_\omega$, correspondingly), the parameter of the $\rho$-$\omega$
mixing $\Pi_{\rho\omega}$  and the width of the $\omega$ meson are
determined in the input file \textbf{cards\_fasterd.dat}. The RPT
parametrization corresponds to the function \textbf{F\_pi\_rpt}.

Inclusion of the $\rho'$ meson modifies the form factor (\ref{formfact}) as
\begin{eqnarray}
F_\pi(q^2)&=&1+\frac{F_V G_V}{f_\pi^2}\frac{q^2}{m_\rho^2}B_\rho^{KS}(q^2)
\Biggl(1-\frac{\Pi_{\rho\omega}}{3m_\omega^2}B_\omega^{KS}(q^2)\Biggr) \\
&+&\frac{F_{V'}
G_{V'}}{f_\pi^2}\frac{q^2}{m_\rho'^2}B_{\rho'}^{KS}(q^2) . \nonumber
\end{eqnarray}

\subsection{"Improved" K\"uhn-Santamaria }
In Ref.~\cite{domin} the pion FF was estimated in the framework of
the dual-QCD$_{N_c\to \infty}$ model. However this model assumes
the resonances to be of the zero width. The authors of Ref.~\cite{kuhn_impr} included the final width
of the $\rho^0$ and the three lowest excited $\rho'$ states  and obtained the following expression for the pion FF
\begin{equation}
F_\pi(q^2)=\sum_{n=0}^3 c_n B_r^{KS}(q^2)+ \sum_{n>4}
c_n\frac{m_n^2}{m_n^2-s} .
\end{equation}
This parametrization for the pion FF is contained in  the function
\textbf{F\_pi\_ks\_new}. The numerical value for the parameters $c_n$ and $m_n$ (the mass of the excited $\rho$ mesons) is taken from Ref.~\cite{kuhn_impr}.

%********************************************************************************************
%         FINAL STATE RADIATION
%********************************************************************************************

\section{Final state radiation models}\label{fsr_model}
Using the underlying symmetry, like gauge invariance,
charge-conjugation symmetry of the final particles and the photon
crossing symmetry, it is possible  to write the FS tensor
$M_F^{(\mu\nu)}$, that describes the $\gamma^*\to\pi^+\pi^-\gamma$
vertex, in terms of three gauge invariant tensors
(see~\cite{our_2005} and Ref. $[23,24]$ therein):
\begin{eqnarray}\label{eqn:fsr}
&&M_{F}^{\mu \nu }(Q,k,l) =\tau _{1}^{\mu \nu }f_{1}+\tau
_{2}^{\mu \nu }f_{2}+\tau _{3}^{\mu \nu }f_{3} ,    \\
&&\tau _{1}^{\mu \nu }=k^{\mu }Q^{\nu }-g^{\mu \nu }k\cdot Q, \;  \; \; \; \; l=p_1-p_2 , \nonumber \\
&&\tau _{2}^{\mu \nu }=k\cdot l(l^{\mu }Q^{\nu }-g^{\mu \nu
}k\cdot
l)+l^{\nu }(k^{\mu }k \cdot l-l^{\mu }k \cdot Q) , \;  \nonumber \\
&&\tau _{3}^{\mu \nu }= s (g^{\mu \nu }k\cdot l-k^{\mu }l^{\nu
})+Q^{\mu }(l^{\nu }k\cdot Q-Q^{\nu }k\cdot l) . \nonumber
\end{eqnarray}
The model dependence comes in only via the
implicit form of the scalar functions $f_i$ (we will call them
structure functions).

Thus the FSR  and interference ($ISR*FSR$) part to the invariant
amplitude squared (\ref{sq_ampl}) is
\begin{eqnarray}\label{mat_fsr}
\overline{|M_{FSR}|^{2}}&=&\frac{1}{s^{2}}\biggl[%
a_{11}|f_{1}|^{2}+2a_{12}\mathrm{Re}(f_{1}f_{2}^{\ast
})+a_{22}|f_{2}|^{2} \nonumber
\\
&+&2a_{23}{Re}(f_{2}f_{3}^{\ast
})+a_{33}|f_{3}|^{2}+2a_{13}{Re}(f_{1}f_{3}^{\ast })\biggr],
\label{fsr}
\end{eqnarray}
and
\begin{eqnarray}\label{mat_ifs}
\mathrm{Re}(\overline{M_{ISR}M_{FSR}^{\ast }})& = & -\frac{1}{4sq^{2}}\biggl[%
A_{1}\mathrm{Re}(F_{\pi }(q^{2})f_{1}^{\ast
})  +A_{2}\mathrm{Re}(F_{\pi }(q^{2})f_{2}^{\ast
}) \\
& + & A_{3}\mathrm{Re}(F_{\pi }(q^{2})f_{3}^{\ast })\biggr]\nonumber .
\end{eqnarray}
The form for the coefficients $a_{ik}$ and $A_i$ can be found in Ref.~\cite{our_2005} Eqs.~(17), (26), correspondingly.

Here is the list of  the  FSR  production mechanisms that are
included in FASTERD:
\begin{eqnarray}\label{fsr_proc}
\text{\hspace{-.3cm}}e^++e^-&\to&\pi^++\pi^-+\gamma
\text{\hspace{3.cm}\textbf{Bremsstrahlung process}}\label{brem}
\\
\text{\hspace{-.3cm}}e^++e^-&\to&\phi\to (f_0;f_0+\sigma)\gamma\to\pi+\pi+\gamma
\label{phi_direct} \text{\hspace{1.5cm}\textbf{$\phi$ direct
decay}}
\\
\text{\hspace{-.3cm}}e^++e^-&\to&(\phi;\omega')\to \rho\pi\to\pi+\pi+\gamma
\label{phi_vmd} \text{\hspace{0.2cm}\textbf{Double resonance
process}} 
\\
\text{\hspace{-.3cm}}e^++e^-&\to&(\rho/\rho')\to \omega\pi^0\to\pi^0+\pi^0+\gamma
\text{ \textbf{Double resonance process}}
\label{rho_vmd}
\end{eqnarray}

Thus the total contribution to the functions $f_i$ introduced in
(\ref{eqn:fsr}) is
\begin{equation}\label{f_i}
f_i=f_i^{(Brem)}+f_i^{(\phi)}+f_i^{(vect)}
\end{equation}

In the next section we present the models describing these
processes. %The presence of the processes (\ref{phi_direct}) and (\ref{phi_vmd} is relevant in  the energy region $s\simeq m_\phi^2$.

\subsection{Final state Bremsstrahlung }
Usually the combined sQED$*$VMD model is assumed for the FS Bremsstrahlung process~\cite{Rr2, phokhara}. In
this case the pions are treated as point-like particles (the sQED
model) and the total FSR amplitude is multiplied by the pion form
factor $F_\pi(s)$, that is estimated in the VMD model.
Unfortunately the sQED$*$VMD model is an approximation that is
valid for relatively soft photons and it can fail for high energy
photons,
 i.e near the $\pi^+\pi^-$ threshold.
In this energy region the contributions to  FSR, beyond the
sQED$*$VMD model, can be important. As mentioned in
Section~\ref{isr},  RPT is supposed to be an appropriate model to
describe the pion-photon interaction in the region about and below
$1$ GeV and this model is used to estimate the contributions beyond
sQED$*$VMD.

Using the sQED$*$VMD model the structure functions $f_i^{(Brem)}$
(see Eq.(\ref{f_i})) are
\begin{eqnarray}
f_1^{sQED}&=&\frac{2k\cdot Q F_\pi(s)}{(k\cdot Q)^2-(k\cdot
l)^2}, \; \; \; f_2^{sQED}=\frac{-2 F_\pi(s)}{(k\cdot
Q)^2-(k\cdot l)^2}, \;\; \\
f_3^{sQED}&=&0 .
\end{eqnarray}
In the framework of RPT the result is
\begin{equation}\label{f}
f^{(Brem)}_{i}=f_{i}^{sQED}+\Delta f^{RPT}_{i},
\end{equation}
where
\begin{eqnarray}\label{d_f}
&&\Delta f^{RPT}_{1}=\frac{F_{V}^{2}-2F_{V}G_{V}}{f_{\pi }^{2}}\biggl(\frac{1}{%
m_{\rho }^{2}}+\frac{1}{m_{\rho }^{2}- s-\mathrm{i}\sqrt{s}\Gamma_\rho(s)}\biggr)  \nonumber \\
&&-\frac{F_{A}^{2}}{f_{\pi }^{2}m_{a}^{2}}\biggl[ 2+\frac{(k\cdot l)^{2}}{%
D(l)D(-l)}+\frac{(s+k\cdot Q)[4m_{a}^{2}-(s+l^{2}+2k\cdot Q)] }{
8D(l)D(-l)}\biggr],  \label{eq:delta-f1} \\
&&\Delta f^{RPT}_{2}=-\frac{F_{A}^{2}}{f_{\pi }^{2}m_{a}^{2}}\frac{%
4m_{a}^{2}-(s+l^{2}+2k\cdot Q)}{8D(l)D(-l)} ,  \label{eq:delta-f2} \\
&&\Delta f^{RPT}_{3}=\frac{F_{A}^{2}}{f_{\pi }^{2}m_{a}^{2}}\frac{k\cdot l}{%
2D(l)D(-l)} ,  D(l)=m_a^2-(s+l^2+2kPQ+4kl)/4 .
\end{eqnarray}

For notations and details of the calculation  we refer the reader
to~\cite{our_2005}. The functions $\Delta f^{RPT}_i$ are calculated by the subroutine \textbf{chpt\_rpt} whereas $f_i^{sQED}$ are evaluated in the
functions \textbf{fsr1\_sqed},  \textbf{fsr2\_sqed}, \textbf{fsr3\_sqed}.

We would like to mention here that the contribution of any model
describing Bremsstrahlung FS process can be  conveniently rewritten
as in Eq.~(\ref{f}) and in the soft photon limit the results should
coincide with the prediction of the sQED$*$VMD model.

\subsection{$\phi$ direct decay}
As it was mentioned in the Introduction, at the DA$\Phi$NE energy
 ($\sqrt{s}=m_\phi $ = 1.01944 \,GeV)
 the final state $\pi\pi\gamma$ can be produced via the intermediate
$\phi$ meson state. In
this section we consider the direct rare decay $\phi\to
\pi\pi\gamma$. As shown in~\cite{graz,gino} this process
affects only the form factor $f_1$ of Eq.~({\ref{eqn:fsr}}):
\begin{equation}\label{phi_ampl}
f_1^{(\phi)}=\frac{g_{\phi\gamma}f_\phi(s)}{s-m_\phi^2+im_\phi\Gamma_\phi}
.
\end{equation}
The $\phi$ direct decay is assumed to proceed through the %$f_0$
intermediate scalar meson state:
 $\phi\to (f_0+\sigma)\gamma\to\pi\pi\gamma$  and its mechanism is
described by a single form factor $f_\phi(s)$.

In the input file the different following models can be chosen to describe
this decay:
\begin{itemize}
\item Non structure model~\cite{bramon}
\item Linear sigma model~\cite{lucio}
\item Chiral unitary approach~\cite{chpt_phi}
\item Achasov kaon loop model~\cite{ach_sol}
\item Achasov kaon loop model with inclusion of the $\sigma$ meson~\cite{achasov}
\item Non structure model that includes both the $f_0$ and $\sigma$ mesons~\cite{gino}
\end{itemize}

The explicit expression for the factor $f_\phi(s)$ for the first three models can be found in~\cite{graz}. For the fourth (``Achasov kaon loop'') model the form factor $f_\phi$ reads:
\begin{eqnarray}
f_\phi^{K^+ K^-}(s) =&& \frac {g_{\phi K^+ K^-} g_{f_0 \pi^+
\pi^-} g_{f_0 K^+ K^-}} {2\pi^2 m^2_{K} (m_{f_0}^2 -s
+Re\Pi_{f_0}(m_{f_0}^2) - \Pi_{f_0}(Q^2))}  \nonumber \\
&& I\left ( \frac {m_\phi^2}{m_K^2},\frac {s}{m_K^2} \right )
e^{i\delta_B(s)} ,
\end{eqnarray}
where $I(.,.)$ is an analytical function~\cite{graz,close} and the phase
$\delta_B(s)=b\sqrt{s-4m_{\pi}^2}$,
$b=75^o/$GeV~\cite{achasov}. The term $Re\Pi_{f_0}(m_{f_0}^2) -
\Pi_{f_0}(s)$ takes into account the finite width corrections to
the $f_0$ propagator~\cite{ach_sol}.

In a refined version of this model which includes the $\sigma$
meson in the intermediate state~\cite{achasov}, the form factor
$f_\phi$ can be written as
\begin{eqnarray*}
f_\phi^{K^+ K^-}(s)&=&\frac{g_{\phi K^+ K^-}}{2\pi^2
m^2_{K}}e^{i(\delta_{\pi\pi}(s)+\delta_{KK}(Q^2))}I\left ( \frac
{m_\phi^2}{m_K^2},\frac {s}{m_K^2}\right ) \\
&& \cdot \sum_{R,R'}g_{RK^+K^-}G^{-1}_{RR'}g_{R'\pi^+\pi^-} ,
\end{eqnarray*}
where $G_{RR'}$ is the matrix of inverse propagators~\cite{achasov}. Such an extension of the model improves the
description of the data at low $m_{\pi\pi}$ and was recently
used by KLOE in the fit of $\phi\to\pi^0\pi^0\gamma$ spectrum~\cite{kloepi0}.

The contribution of the $\phi$ direct decay is calculated in the
subroutines \textbf{fph1\_bcg, fph1\_ln, fph1\_cpt, fph1\_a4q, fph1\_a4qs, fph1\_pac} according to the corresponding model.

\subsection{Double resonance contribution}\label{fsr_model_2res}
Another mechanism producing the final $\pi\pi\gamma$ state is
reported in Eqs.~(\ref{phi_vmd}), (\ref{rho_vmd}). 

First we consider the $\phi\to\rho\pi$ mechanism. In this case the  $\phi$
meson decays in ($\rho^\pm \pi^\mp$) for the charged channel and
in ($\rho^0\pi^0$) for the neutral one and then $\rho
\to\pi\gamma$.
%The implicit value for the functions $f_i^{VMD}$ for this decay
%can be found in Ref. \cite{gino}.
The corresponding contribution to the functions $f_i^{(vect)}$ of
Eq.~(\ref{f_i}) is
\begin{eqnarray}
\label{eq:delta-f1_pi_pi} f_{1}^{\phi\rho} &=&
\frac{C(s)}{16\pi\alpha}\bigl[ ({k\cdot
Q +l^2}) \bigl(D_{\rho}(R^2_{+}) + D_{\rho}(R^2_{-}) \bigr)+ 2 k
\cdot l \bigl( D_{\rho}(R^2_{+}) -
D_{\rho}(R^2_{-}) \bigr) \bigr] , \nonumber \\
f_{2}^{\phi\rho} &= &
-\frac{C(s)}{16\pi\alpha} \bigl[
D_{\rho} (R^2_{+})+ D_{\rho} (R^2_{-}) \bigr]
, \nonumber \\
f_{3}^{\phi\rho} &=&
\frac{C(s)}{16\pi\alpha} \bigl[
D_{\rho}(R^2_{+}) - D_{\rho}(R^2_{-}) \bigr], \nonumber
\end{eqnarray}
where
\begin{equation}\label{coeff_vmd}
C(s)=\frac{\sqrt{4\pi\alpha}\;g^\rho_{\pi\gamma}g^\phi_{\rho\pi} \; F_\phi\; s}{m_\phi^2-s-im_\phi
\Gamma_\phi}
\end{equation}
and $D_{\rho}(R)=m_\rho^2-R-i m_\rho \sqrt{R}\Gamma_\rho(R)$,
$R_+=(k+p_1)^2$, $R_-=(k+p_2)^2$. The quantities
$g^\phi_{\rho\pi}$, $g^\rho_{\pi\gamma}$ are the coupling
constants determining respectively the $\phi\to\rho\pi$ and $\rho\to\pi\gamma$
vertexes correspondingly,
$F_\phi=\sqrt{\displaystyle\frac{\mathstrut 3\Gamma(\phi\to
e^+e^-)}{4\pi\alpha^2 m_\phi}}$.

To make connection with  the KLOE analysis \cite{kloepi0} we add also the
phase of the $\omega$-$\phi$ meson mixing $\beta_{\omega\phi}$ and
the constant factor $\Pi_\rho^{VMD}$~\footnote{Including $\Pi_\rho^{VMD}$,
in our opinion, rescales the
constant $g^\phi_{\rho\pi}$ that cannot be directly determined
from any experimental decay width}.

Thus the coefficient (\ref{coeff_vmd}) is rewritten as
\begin{equation}
C(s)=\frac{\sqrt{4\pi\alpha}\; g^\rho_{\pi\gamma}\; g^\phi_{\rho\pi}\; F_\phi\; s}{m_\phi^2-s-im_\phi
\Gamma_\phi}\Pi_\rho^{VMD} e^{i \beta_{\omega\phi}} .%e^{i\delta_\rho}}
\end{equation}

In the energy region of DA$\Phi$NE also the tail of the excited
$\omega$  meson can play a role:
$\gamma^*\to\omega'\to\rho\pi$. The explicit form of this
contribution is written similar to (\ref{eq:delta-f1_pi_pi}) and is:
\begin{equation}\label{c_omega}
C^\omega_{\rho\pi}=\frac{\sqrt{4\pi\alpha}\;g^\rho_{\pi\gamma}\;g^{\omega'}_{\rho\pi}\;F_{\omega'}\;
s}{m_{\omega'}^2-s-im_{\omega'} \Gamma_{\omega}} .
\end{equation} 

In the considered energy region
Eq.~(\ref{c_omega}) can be approximated by a complex constant
$C^\omega_{\rho\pi}$ whose the numerical value  is taken
from the fit of $\pi^0\pi^0\gamma$ spectrum~\cite{kloepi0}\footnote{
Also if we suppose that the direct decay
$\gamma^*\to\rho\pi\to\pi\pi\gamma$ takes place it contributes to
$C^\omega_{\rho\pi}$.}.

Therefore we have
\begin{equation}\label{coeff_c_final}
C(s)=\frac{\sqrt{4\pi\alpha}\;g^\rho_{\pi\gamma} g^\phi_{\rho\pi}\;F_\phi\;s}{m_\phi^2-s-im_\phi
\Gamma_\phi}\Pi_\rho^{VMD}
e^{i \beta_{\omega\phi}}+C^\omega_{\rho\pi} .
\end{equation}

For the neutral case the
$\gamma^*\to(\rho/\rho')\to\omega\pi^0\to\pi^0\pi^0\gamma$
reaction contributes as well. 
As for the $\omega'\to\rho\pi$ channel we
replace the explicit expression
\begin{equation}\label{c_rho}
C^\rho_{\omega\pi}=\frac{\sqrt{4\pi\alpha}\;g^\omega_{\pi\gamma}\;g^{\rho'}_{\omega\pi}\;F_{\rho'}\;
s}{m_{\rho'}^2-s-im_{\rho'} \Gamma_{\rho'}}+
\frac{\sqrt{4\pi\alpha}\;g^\omega_{\pi\gamma}\;g^{\rho}_{\omega\pi}\;F_{\rho}\;
s}{m_{\rho}^2-s-im_{\rho} \Gamma_{\rho}}
\end{equation}
by a complex constant $C^\rho_{\omega\pi}$, which  is again taken from the fit. Then the contribution of the $\gamma^*\to(\rho/\rho')\to\omega\pi^0\to\pi^0\pi^0\gamma$ mechanism to the functions $f_i^{(vect)}$ is
\begin{eqnarray}
\label{omega_pi}
f_{1}^{\rho\omega} &=&
\frac{C^\rho_{\omega\pi}}{16\pi\alpha}\bigl[ ({k\cdot Q +l^2})
\bigl(D_{\omega}(R^2_{+}) + D_{\omega}(R^2_{-}) \bigr)+ 2 k \cdot
l \bigl( D_{\omega}(R^2_{+}) -
D_{\omega}(R^2_{-}) \bigr) \bigr] , \nonumber \\
f_{2}^{\rho\omega} &= & -\frac{C^\rho_{\omega\pi}}{16\pi\alpha}
\bigl[ D_{\omega} (R^2_{+})+ D_{\omega} (R^2_{-}) \bigr]
, \nonumber \\
f_{3}^{\rho\omega} &=& \frac{C^\rho_{\omega\pi}}{16\pi\alpha} \bigl[
D_{\omega}(R^2_{+}) - D_{\omega}(R^2_{-}) \bigr] . \nonumber
\end{eqnarray}

Finally the total contribution  is
\begin{equation}\label{f_i_vect}
f_i^{(vect)}=f_i^{(\phi\rho)} e^{i\delta_\rho}+c f_i^{(\rho\omega)} ,
\end{equation}
where $c=1$ for the neutral final state and $c=0$ for the charged
one. We include also an additional phase between the double
resonance and $\phi$  direct contributions ($\delta_\rho$).

The subroutine \textbf{rhotopig} calculates the contribution for the double resonance mechanism.

%******************************************************************************
%                 COMPARISON WITH ANALYTICAL PREDICTION. DISCUSSION
%******************************************************************************

\section{Comparison of MC results with analytical prediction. \textbf{Discussion of the results}}\label{comparison}

For the full angular range
$0^\circ<\theta_\gamma;\theta_\pi<180^\circ$ it is possible to
obtain the analytical expression of the cross section for the
processes (\ref{process_charged}) and (\ref{process_neutral}).
Here we present the results for the charged final state\footnote{For the
$\pi^0\pi^0\gamma$ case
the contribution both from the $\phi$ direct decay and the double resonance vector mechanism is a half of
the corresponding one for the  $\pi^+\pi^-\gamma$ final
state (except for the $\gamma^*\to\rho\to\omega\pi$ case that
appears only for the $\pi^0\pi^0\gamma$ FS).}.

The ISR cross section is
\begin{equation}\label{isr_anal}
\frac{d\sigma^{(ISR)}}{dq^2}=\frac{\alpha(s^2+q^4)}{3s^2q^2(s-q^2)}|F_\pi(q^2)|^2
\Biggl(1-\frac{4m_\pi^2}{q^2}\Biggr)^{3/2}(L-1) , \; \; \;
L=\ln{\frac{s}{m_e^2}} ,
\end{equation}
whereas for the FSR process we have
\begin{equation}\label{fsr_anal}
\frac{d\sigma^{(FSR)}}{dq^2}=\frac{\alpha^3(s-q^2)}{24\pi
s^3}(2h_1-s h_2) ,
\end{equation}
where the form of the functions $h_i$ and the equations to
calculate them using the functions $f_i$ can be found in
Ref.~\cite{our_2005}. For example, in the case of sQED
\begin{equation}
2h_1-s h_2=\frac{16\pi\xi
|F_\pi(s)|^2}{(kQ)^2}\Biggl[(kQ)^2+\Biggl(\frac{s}{4}-m_\pi^2\Biggr)(-q^2+(q^2-2m_\pi^2)L_1)\Biggr]
\end{equation}
with $\xi=\sqrt{1-\displaystyle\frac{\mathstrut 4m_\pi^2}{q^2}}$
and $L_1=\displaystyle\frac{\mathstrut
1}{\xi}\ln{\frac{1+\xi}{1-\xi}}$.

\begin{figure}
\begin{center}
\par
\parbox{1.0\textwidth}{\hspace{-0.5cm}
%\vspace{5.cm}
\includegraphics[width=0.5\textwidth,height=0.45\textwidth]{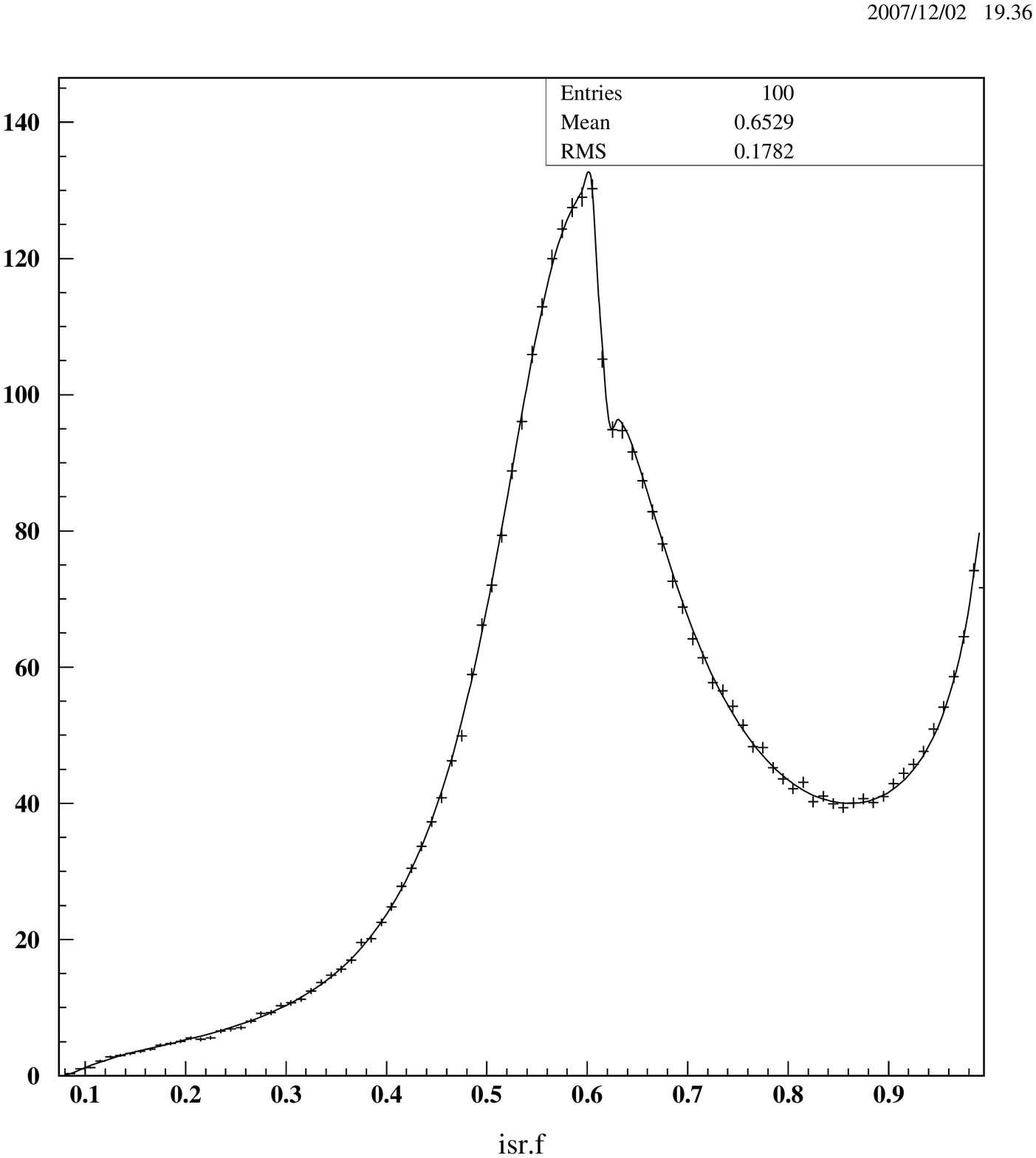}
\includegraphics[width=0.5\textwidth,height=0.45\textwidth]{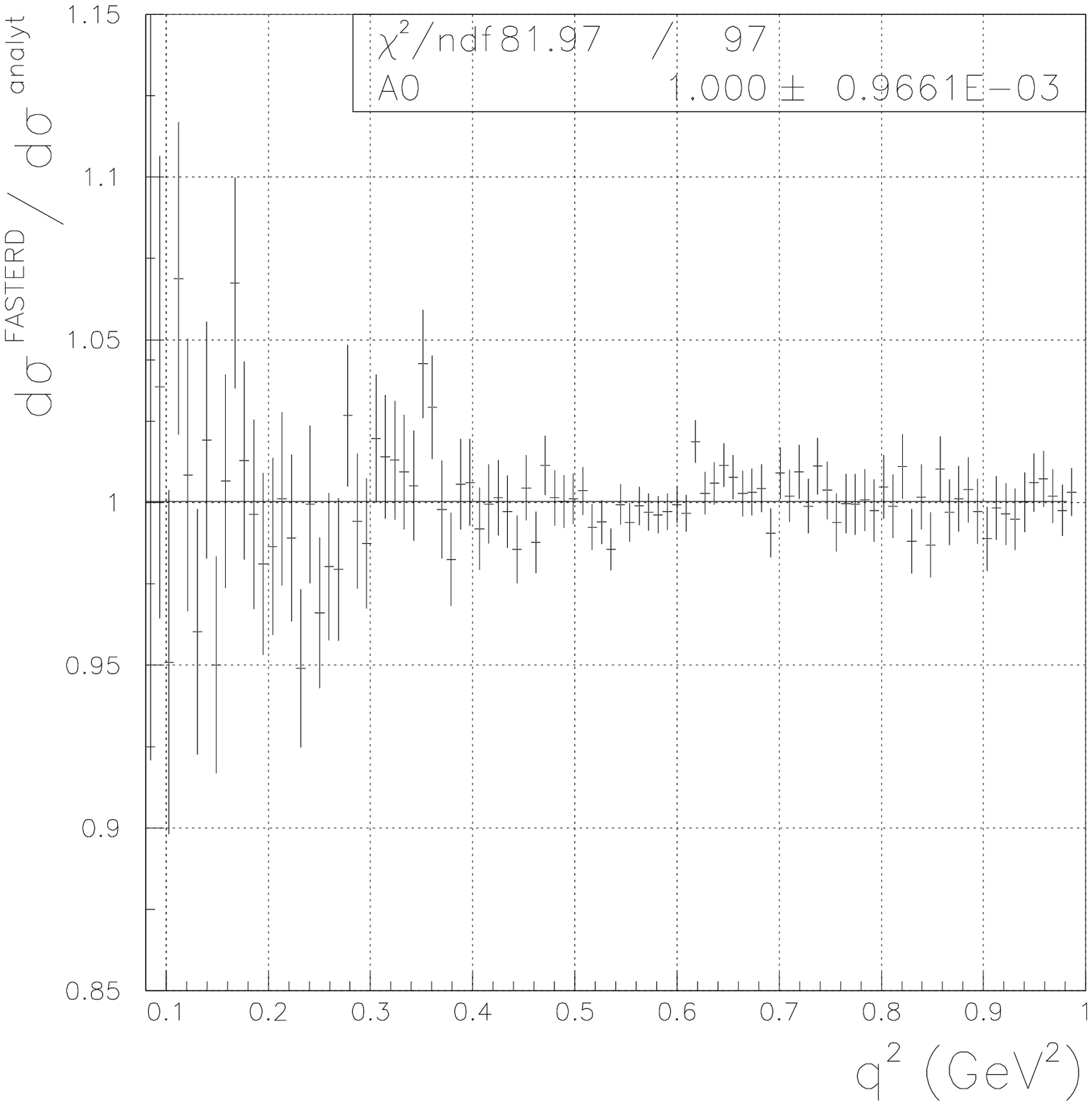}}
\caption{Comparison of the analytical and numerical results for
ISR: $0^\cdot<\theta_\pi<180^\cdot$,
$0^\cdot<\theta_\gamma<180^\cdot$. Left: the solid line corresponds to the analytical result whereas the points with errors have been obtained by FASTERD. Right: the ratio between the analytical results and numerical ones is fitted by the constant $A_0$.}\label{fig_compar_isr}
\end{center}
\end{figure}

\begin{figure}
\begin{center}
\par
\parbox{1.0\textwidth}{\hspace{-0.5cm}
\includegraphics[width=0.5\textwidth,height=0.45\textwidth]{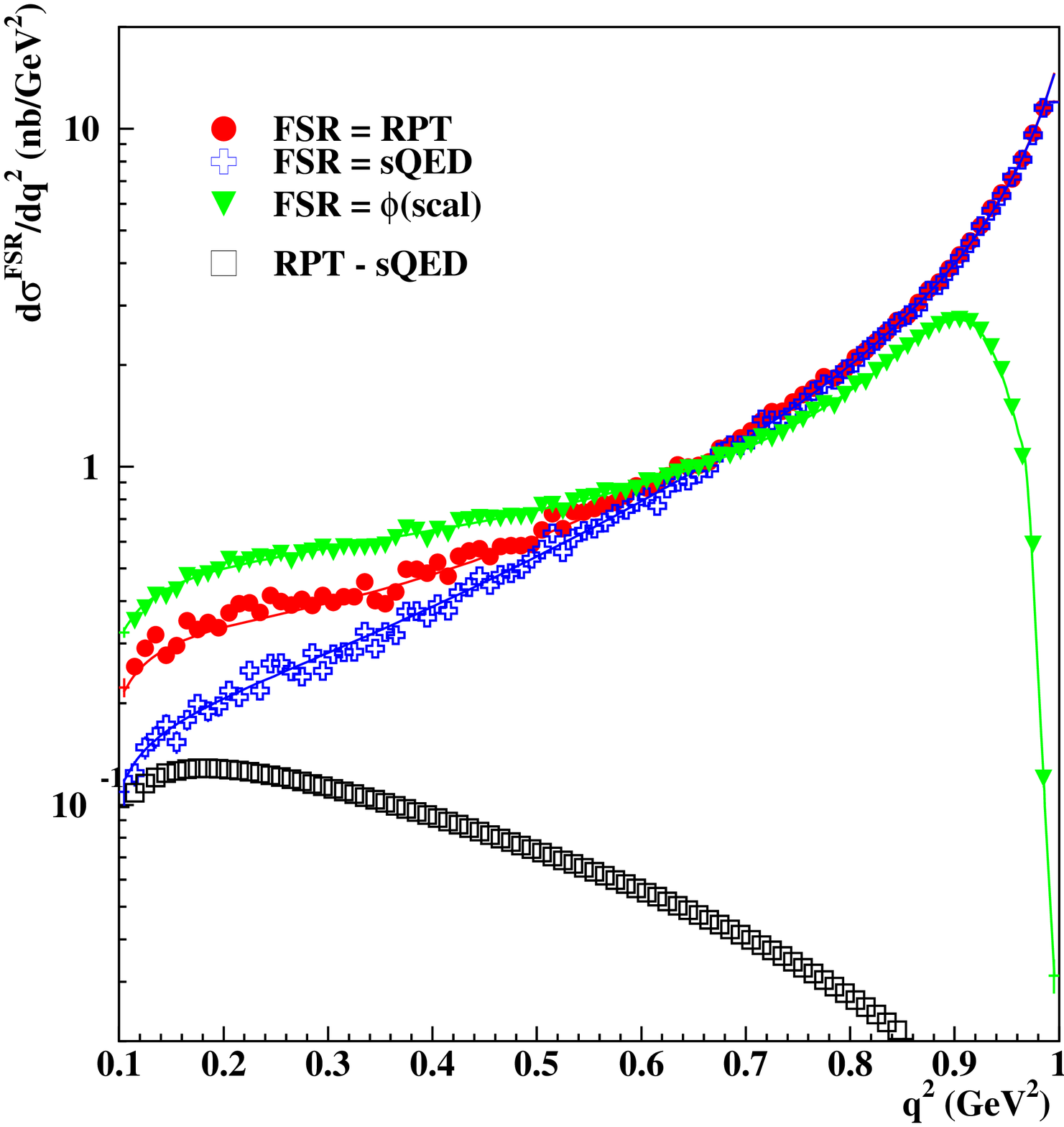}
\includegraphics[width=0.5\textwidth,height=0.45\textwidth]{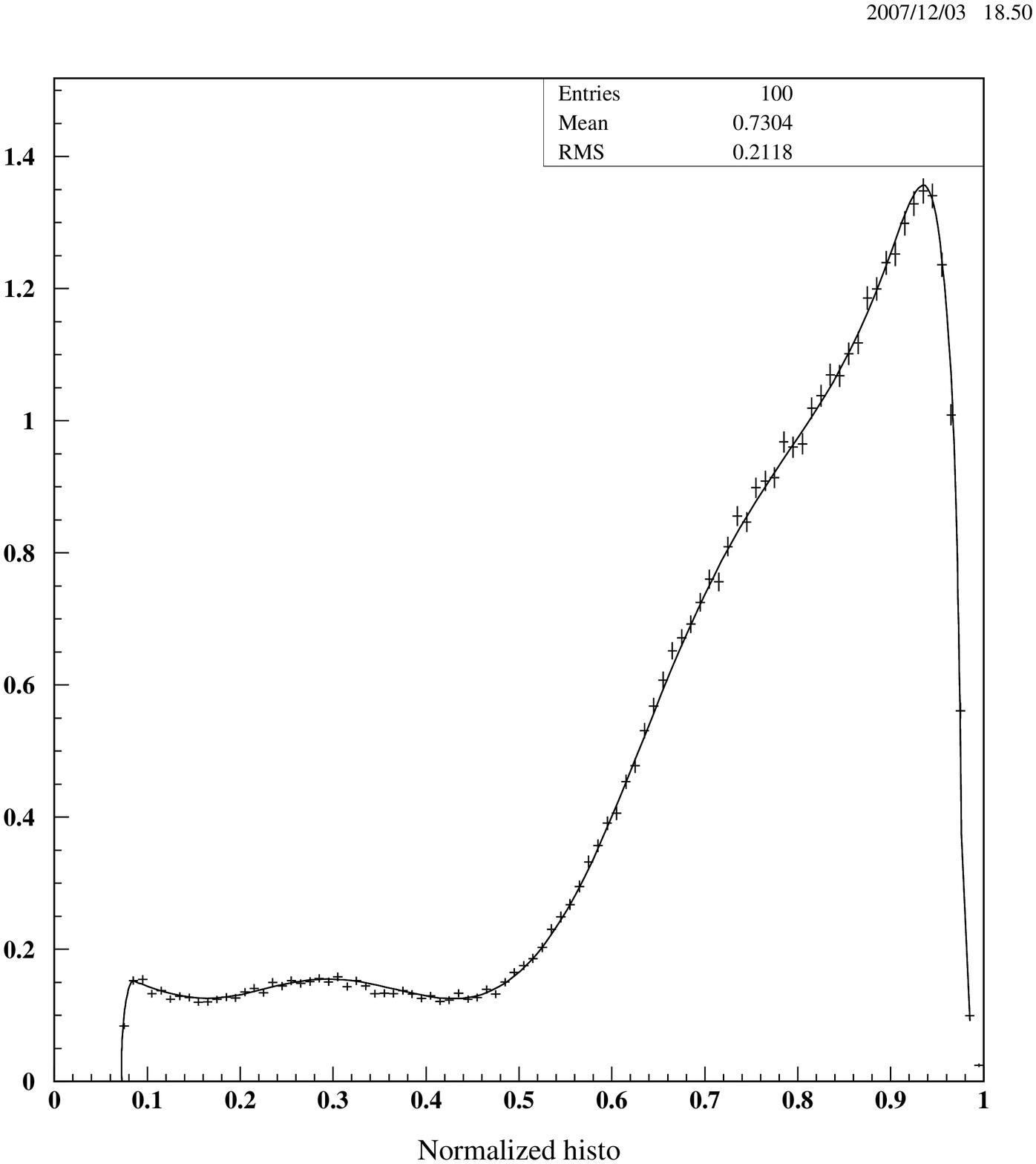}}
\caption{Comparison of the analytical and numerical results for
different contributions to FSR: $0^\cdot<\theta_\pi<180^\cdot$,
$0^\cdot<\theta_\gamma<180^\cdot$. For the $\phi$ direct decay the
kaon loop model with $f_0+\sigma$ is chosen whereas the RPT
parametrization is chosen for the pion FF. The solid lines correspond to the analytical result whereas the points (triangles, squares) have been obtained by FASTERD.}\label{fig_compar_fsr}
\end{center}
\end{figure}

In Figs.~\ref{fig_compar_isr}  and \ref{fig_compar_fsr} we compare the analytical predictions
(\ref{isr_anal}) and (\ref{fsr_anal}) with the numerical results
obtained by FASTERD for the cross section of the reaction
$e^+e^-\to\pi^+\pi^-\gamma$. The ratio between the analytical and numerical results is fitted by the constant  $A_0$. We have good $\chi^2$ and the value of $A_0$ compatible with one.  
%The agreement is excellent.
%\textbf{(it is better than $0.5\%$)}.

\begin{figure}
\begin{center}
\par
\parbox{1.0\textwidth}{\hspace{-0.5cm}
%\vspace{5.cm}
\includegraphics[width=0.5\textwidth,height=0.6\textwidth]{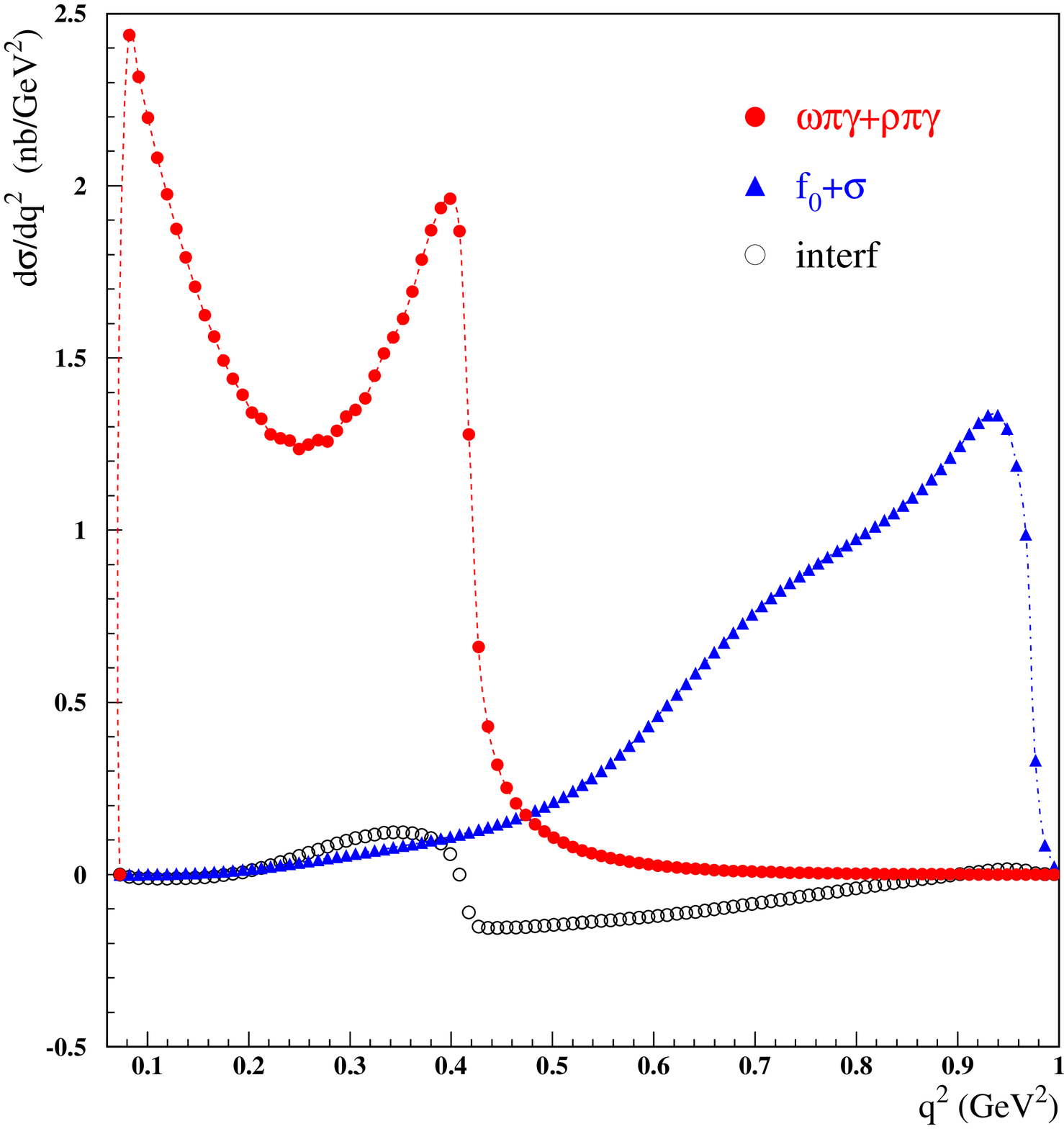}
\includegraphics[width=0.5\textwidth,height=0.6\textwidth]{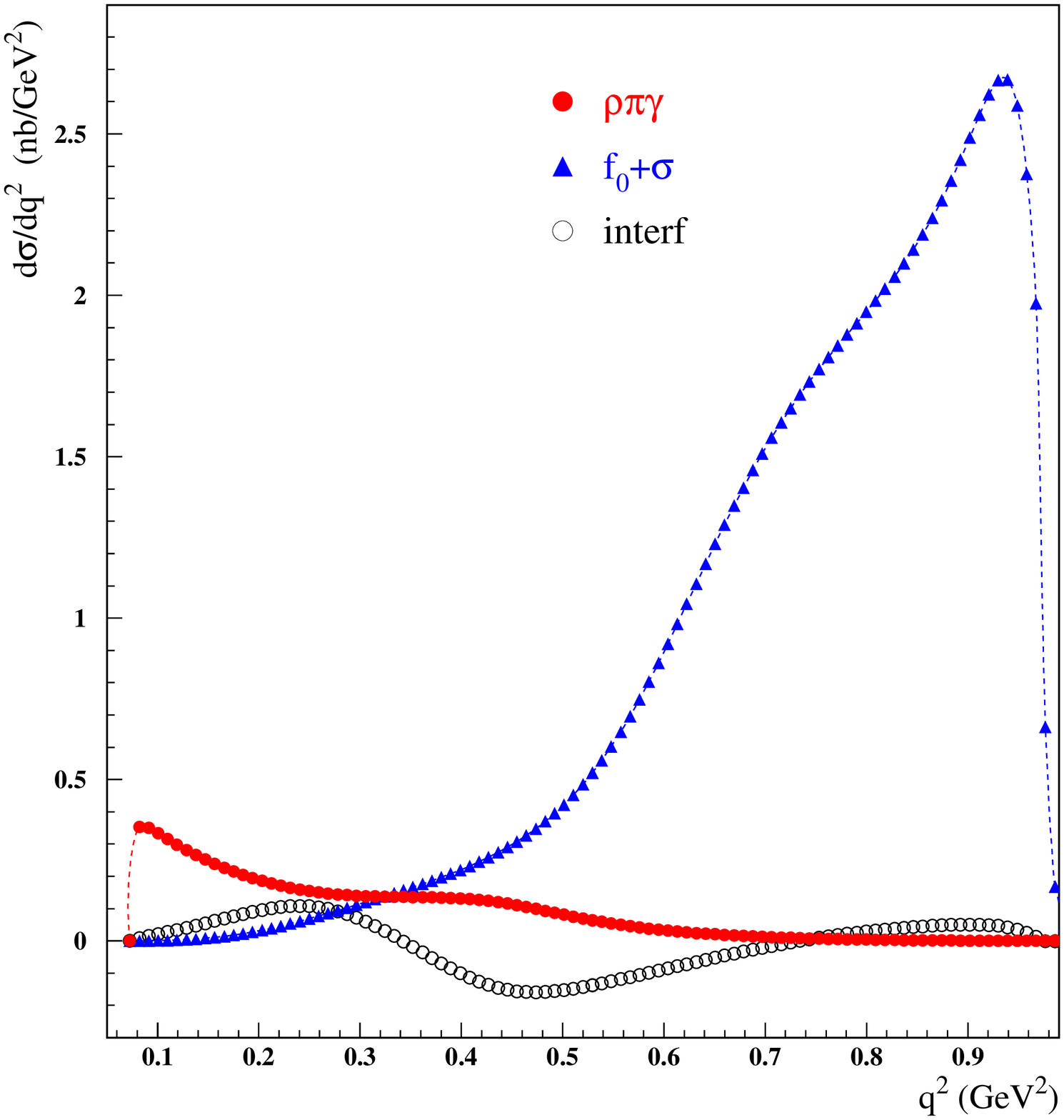}
}
%\hspace{1.cm}
\vspace{-1.cm} \caption{Different contributions to FSR for the
$\pi^0\pi^0\gamma$ (left) and $\pi^+\pi^-\gamma$ (right) final
states: $0^\cdot<\theta_\pi<180^\cdot$,
$0^\cdot<\theta_\gamma<180^\cdot$. The filled circles  are for  the double
resonance contribution, the triangles correspond to the
$(f_0+\sigma)\gamma$ mechanism and the empty circles correspond to  the
interference term between the $(f_0+\sigma)\gamma$ and the double resonance contributions.}\label{fig_double_vector}
\end{center}
\end{figure}

An important feature of FASTERD is a possibility to
estimate the double resonance contribution. At the first sight in
the energy region about $s\approx m_\phi^2$ it is enough to
include only the $\gamma^*\to\phi\to\rho\pi\to\pi\pi\gamma$
mechanism. However, at $s=m_\phi^2$ our simulation gives the
$\sigma(\phi\to ((f_0+\sigma)\gamma+\rho^0\pi^0) \to
\pi^0\pi^0\gamma) = 0.451\pm 0.001$ nb whereas $\sigma((\rho/\rho')\to
\omega\pi^0 \to \pi^0\pi^0\gamma) = 0.529\pm 0.012$ nb (in agreement with
KLOE results~\cite{giovannella_pi0}). In the case
of the $\pi^+\pi^-\gamma$ the double resonance mechanism (only the
$\phi\to\rho^\pm\pi^\mp$ contributes) does not give a large contribution in the  high $q^2$ region whereas it gives visible
the cross section in the low $q^2$ region (Fig.~\ref{fig_double_vector}, right).
In the current version FASTERD allows to simulate the
processes (\ref{process_charged}) and (\ref{process_neutral}) for
$s\approx m_\phi^2$. To simulate the processes for lower (upto the threshold) and
higher (say upto 1.2 GeV$^2$) energies the exact value of the constants $C^\omega_{\rho\pi}$ and
$C^\rho_{\omega\pi}$ (Eq.~(\ref{c_omega}) and Eq.~(\ref{c_rho})) has to be used.

\begin{figure}
\begin{center}
\par
%%\parbox{1.0\textwidth}{\hspace{-0.5cm}
%\vspace{5.cm}
\includegraphics[width=0.5\textwidth,height=0.5\textwidth]{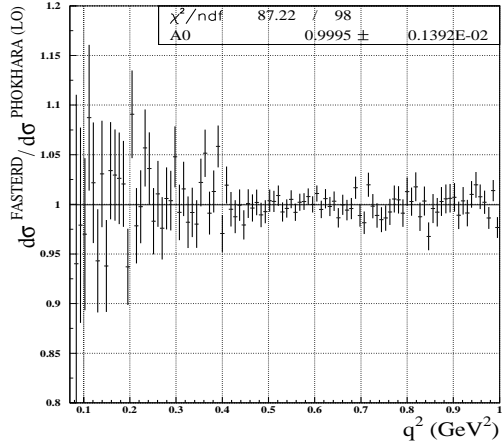}
%%\includegraphics[width=0.5\textwidth,height=0.5\textwidth]{compar_fasterd_phokhara_isr_ksdual_nlo_0_180.eps}}
%\hspace{1.cm}
%%\vspace{-1.5cm}
\caption{Comparison of the numerical results for ISR simulated by
FASTERD and PHOKHARA : $0^\cdot<\theta_\pi<180^\cdot$,
$0^\cdot<\theta_\gamma<180^\cdot$. The "improved"
K\"uhn-Santamaria parametrization is chosen for the pion
FF. The ratio between the numerical results obtained by FASTERD and PHOKHARA is fitted by the constant $A_0$.}\label{fig_compar_phokhara}
\end{center}
\end{figure}

We compare the ISR spectrum  with the results from PHOKHARA~\cite{phokhara}. The 'improved' K\"uhn-Santamaria parametrization was chosen for the pion form factor and PHOKHARA was run at the leading order approximation (only one photon is radiated). The results of
comparison are presented in Fig.~\ref{fig_compar_phokhara}. There the ratio between FASTERD and PHOKHARA results is fitted by the constant $A_0$. As one
can see the PHOKHARA result coincides with the FASTERD prediction. 
As it was discussed in Ref.~\cite{phokhara}
the radiative corrections due to an  additional photon emitted by the leptons are relevant and we are planning to include them in the code. 
An additional photon can be
emitted either from initial state or from final state. Inclusion
of the second photon emitted from the initial state radiation is
a technical task and can be obtained by replacing the leptonic tensor (it has been calculated
already~\cite{phokhara, our_2005}) and substitution $k\to k'$, $Q\to Q'= p_+-p_--k'$ in Eq.~(\ref{eqn:fsr}),
where $k'$ is the energy of the additional photon emitted from the
initial state. Also the one photon phase space has to be replaced
to the two photon phase space. For the second photon from the
final state the situation is more difficult as there is not yet a
model-independent calculation for the final state tensors with two
photons.

%**********************************************************************************
%          SPECTRUM AND GENERAL ALGORITH
%**********************************************************************************

\section{The calculation of the spectrum. General algorithm}\label{subroutine}

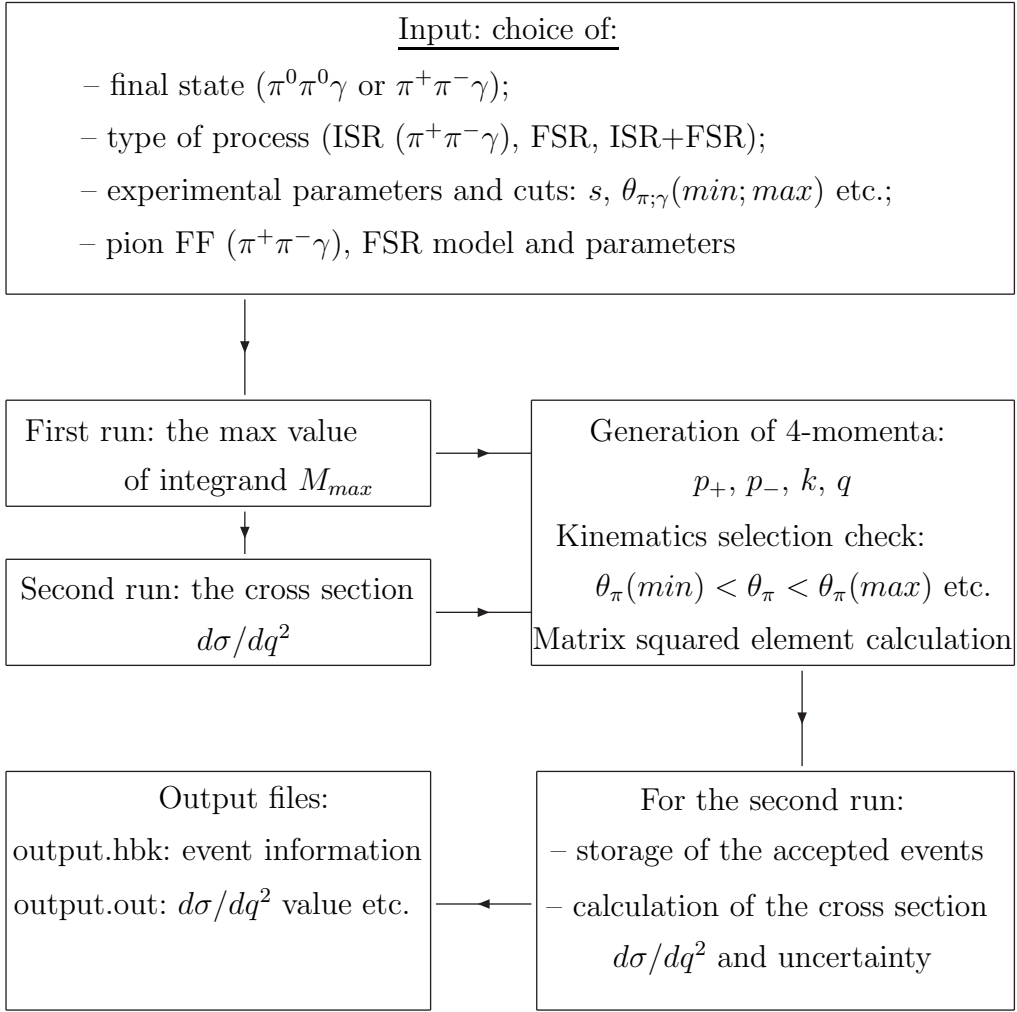
\begin{figure*}
\vspace*{.5cm}
\begin{picture}(600,420)(10,30)

\Line(10,460)(390,460) \Line(10,350)(390,350)
\Line(10,460)(10,350) \Line(390,460)(390,350)

\Text(200,450)[]{\underline{Input: choice of:}}
\Text(122,430)[]{-- final state ($\pi^0\pi^0\gamma$ or
$\pi^+\pi^-\gamma$);     }

\Text(168,410)[]{-- type of process (ISR ($\pi^+\pi^-\gamma$),
FSR, ISR$+$FSR);} \Text(192,390)[]{-- experimental parameters and
cuts: $s$, $\theta_{\pi;\gamma}(min;max)$ etc.;}
\Text(162,370)[]{-- pion FF ($\pi^+\pi^-\gamma$), FSR model
and parameters}

\ArrowLine(100,348)(100,312)

\Line(10,310)(170,310) \Line(10,210)(170,210)
\Line(208,310)(390,310) \Line(208,210)(390,210)
\Text(80,300)[]{First run: the max value}
\Text(100,280)[]{ of integrand $M_{max}$}
\Line(10,270)(170,270) \Line(10,250)(170,250)
\Text(90,240)[]{Second run: the cross section}
\Text(100,220)[]{$d\sigma/dq^2$}

\Line(10,310)(10,270) \Line(170,310)(170,270)
\Line(10,250)(10,210) \Line(170,250)(170,210)

\Line(208,310)(208,210) \Line(390,310)(390,210)
\ArrowLine(172,290)(208,290)
\ArrowLine(172,230)(208,230)

\Text(300,300)[]{Generation  of 4-momenta: }
\Text(300,280)[]{$p_+$, $p_-$, $k$,
$q$}
\Text(290,260)[]{Kinematics selection check: }
\Text(310,240)[]{$\theta_\pi(min)<\theta_\pi<\theta_\pi(max)$
etc. }
\Text(300,220)[]{Matrix squared element calculation}
\ArrowLine(100,268)(100,252)

\Line(210,170)(390,170) \Line(210,80)(390,80)
\Line(210,170)(210,80) \Line(390,170)(390,80)
\Text(300,160)[]{For the second run:}
\Text(300,140)[]{-- storage of the accepted events }
%\Text(200,120)[]{angles and 4-momentum all particles}
\Text(300,120)[]{-- calculation of the cross section }
\Text(300,100)[]{$d\sigma/dq^2$ and uncertainty}

\Line(10,170)(170,170) \Line(10,80)(170,80)
\Line(10,170)(10,80) \Line(170,170)(170,80)
\Text(100,160)[]{Output files:}
\Text(90,140)[]{output.hbk: event information}
\Text(90,120)[]{output.out: $d\sigma/dq^2$ value etc. }
%\Line(390,60)(390,0)
\ArrowLine(208,120)(172,120)
\ArrowLine(310,208)(310,172)
%\Line(10,-20)(390,-20) \Line(10,-85)(390,-85)
%\Line(10,-20)(10,-85) \Line(390,-20)(390,-85)
%\Text(200,-45)[]{ \ \hfill }

\end{picture}
\vspace*{-2.cm}
\caption{The logical structure of FASTERD}\label{fig_algorithm}
%\noindent {\it Figure 1: } \vspace*{1.2cm}
\end{figure*}

The logical structure of FASTERD is presented in Fig.~\ref{fig_algorithm}.

Input parameters are read from the file
\textbf{cards\_fasterd.dat}  that is modified by the  user (see
Appendices A, B for the description of the input file). Here it is
possible to choose the name of the output files.  The main part of
the program runs twice. On the first step, in order to estimate
the maximum value of the integrand ($M_{max}$) some trial events
(NMPT) are generated.
 On the second step (NEVE) events are produced and the acceptance-rejection method
is applied to evaluate  the cross section $d\sigma/dq^2$.
The generator RANMAR~\cite{cern_lib} is used to generate
the random numbers. The variables ($q^2$, $\cos\theta_\gamma$,
$\cos\phi_\gamma$) are generated in the laboratory frame ($e^+e^-$
center of mass system), the variables ($\cos\theta_\pi^{\pm}$ and
$cos\phi_\pi^{\pm}$) are generated in the pion mass reference
system. They are calculated in the subroutines \textbf{qquadrat}
and \textbf{photonangles}. The subroutine \textbf{vectorb}
transforms the azimuthal and polar angles of pions in the
laboratory system. The subroutine \textbf{rejection} checks the
restrictions imposed by the input file. The matrix squared element
is produced inside the  function \textbf{mat}. Using the value
$mat$ returned by this function the differential cross section
$inte$ is computed as $inte=mat*num*jac\_pion*jac\_phot$, where
$num$ is a kinematic factor for the cross section, $jac\_pion$ and
$jac\_phot$ are the replacement factors  (they are calculated in
the subroutines \textbf{qquadrat} and \textbf{photonangles}). The
information about the accepted events is stored in the file
\textbf{output.hbk}. At the end of the generation, the cross
section is computed by the ratio of accepted to generated events,
and the error is estimated.
 The information about the number of generated
and accepted event as well as the value of the cross section  and its error is written in a output file.

Function \textbf{mat}. The input for \textbf{mat} is the 4-momenta
of all particles in the laboratory frame and the type of
process (ISR, FSR or both ISR$+$FSR). According to the type of
the process either the subroutine \textbf{matr\_isr} or
\textbf{matr\_fsr} or \textbf{matr\_int} are called. They compute the Eqs.~(\ref{mat_isr}),
 (\ref{mat_fsr}), (\ref{mat_ifs}) presented in this paper.

Subroutine \textbf{qquadrat}. The following probe function is used to generate the variables $q^2$,
$\cos\theta_\pi$ and $cos\phi_\pi$ (we remind that the angles of the pions are generated in the pion reference system)
\begin{eqnarray*}
f(q^2,\cos\theta_\pi)&=&\frac{1}{s(s-q^2)}+\frac{1}{(q^2-m_\rho^2)^2+\Gamma_\rho^2 m_\rho^2}+ \\
&&\frac{1}{s(s-q^2)}\Biggl(\frac{1}{1-\sqrt{1-\frac{4m_\pi^2}{s}}\cos\theta_\pi}+
\frac{1}{1+\sqrt{1-\frac{4m_\pi^2}{s}}\cos\theta_\pi}\Biggr) .
\end{eqnarray*}
The polar pion angle is generated uniformly.

It results in  the following expression for the $jac\_pion$
\begin{equation}
jac\_pion=\frac{2\pi V}{f(q^2,\cos\theta_\pi)} ,
\end{equation}
where
\begin{eqnarray*}
V&=&-\frac{2}{s}\ln\frac{s-q_{max}^2}{s-q_{min}^2}+
\frac{1}{\Gamma_\rho m_\rho}\Biggl(arctan\frac{q_{max}^2-m_\rho^2}{\Gamma_\rho m_\rho}-arctan\frac{q_{min}^2-m_\rho^2}{\Gamma_\rho m_\rho}\Biggr)\\
&-&
\frac{1}{s(1-\sqrt{1-\frac{4m_\pi^2}{s}})^2\sqrt{1-\frac{4m_\pi^2}{s}}}
\ln\frac{1+\sqrt{1-\frac{4m_\pi^2}{s}}}{1-\sqrt{1-\frac{4m_\pi^2}{s}}} .
\end{eqnarray*}

%Subroutine \textbf{}

 \vspace{1cm}

\section*{Acknowledgements}
 We are grateful to H. Czy\.z, S. Eidelman, F. Jegerlehner, H. K\"uhn, and G. Rodrigo, A. Sibidanov and the members of the KLOE collaboration
 for many useful discussions.
 G.P. acknowledges  support from EU-CT2002-311 Euridice contract. This work has been supported by the grant INTAS/05-1000008-8328.

\begin{appendix}
\section*{Appendix A. Input file}\label{App_A}
\setcounter{equation}{0}
\def\theequation{A.\arabic{equation}}

The main parameters are
set in \textbf{cards\_fasterd.dat} input file. It defines the cuts used in the generation, and the parameters needed for the for the FSR model and the pion FF.
%generation, the parameters for
%the experimental setting and, at last,   the numerical value of
%the parameters for the FSR model and the pion FF.

The following variables are included in
\textbf{cards\_fasterd.dat} file:
\begin{itemize}
\item NEVE - number of the generated events to be generated
%to estimate the cross section
\item NPTM - number of events to find the maximum of the cross section.
\item NRND - random seed
%\item CLRD - whether the program should include the radiative corrections calculted by the subroutine \textit{radiation}.
\item FSKIND - kind of  final state (FSKIND$=$0 for $\pi^0\pi^0$ and FSKIND$=$1 for $\pi^+\pi^-$)
\item KIND - two parameters which allows to choose the kind of radiation: (ISR correspond to (1 0), FSR to (0 1) and ISR$+$FSR to (1 1))
\item FPKIND - defines the pion FF parametrization
%(FPKIN$0$ for KS, FPKIND$=1$ for GS, FPKIND$=3$ for the pion FF in the framework of RPT and FPKIND$=4$ for ``improved'' KS)
($1$ for KS, $2$ for GS, $3$ for the pion FF in the framework of RPT and $4$ for ``improved'' KS)
\item RHOKIND - choice of FSR contribution
\item BREMKIND - choice of Bremsstrahlung model
\item f0KIND - choice of the model for the $\phi$ direct decay
\item RESPIKIND - allows to choose the
 intermediate states ($\rho\pi$ and/or $\omega\pi^0$) for
the double resonance contribution\footnote{The $\omega\pi$ intermediate state is produced only for the
$\pi^0\pi^0$ final state}.
\item HOUT - defines the name of the output file \textbf{output.hbk} where information of the generated particles (4-momenta) is inserted.
\end{itemize}

The parameters that define the experimental cuts are
\begin{itemize}
\item ENES - the squared energy of the initial particles
%\item EMIN - the minimal energy of the radiated photon
%\item QMIN - the minimal value of the pion invariant mass squared
\item GMIN - the minimal energy of the radiated photon
\item GMAX - the maximal energy of the radiated photon
\item ACUT -  the minimal and maximal azimuthal angles of photon and pions
\end{itemize}

The value of the constants for the pion FF are
defined in the arrays \textit{FP} and \textit{FPIHIGH}. The remaining set of arrays
\textit{F0par,F0fase,F0nonstr,VMDPAR} defines the numerical values
for the $\phi$ direct decay and the
double resonance contribution.

\end{appendix}

\begin{appendix}
\section*{Appendix B. Example of input  file}\label{App_C}
\setcounter{equation}{0}
\def\theequation{B.\arabic{equation}}

As an example, the input file \textbf{cards\_fasterd.dat} is set
%Here we give, as an example,
%the input  file
to generate the  $e^+e^-\to\pi^+\pi^-\gamma$ process where:
\begin{itemize}
\item the photon is emitted both from IS and FS;
\item  the RPT parametrization is chosen for the FF;
\item  FSR is given by sQED Bremsstrahlung and $\phi\to(f_0+\sigma)$  decay;
\item the experimental cuts are: $50^\circ<\theta_{\pi,\gamma}<130^\circ$
\item the output file is named \textbf{isr\_rpt\_fsr\_sqed\_f0$+$sig\_50\_130\_2e6.hbk}
\end{itemize}

\begin{verbatim}
C* --- number of events to be processed
NEVE      1000000
C* --- random seed
NRND      9513463
C* --- number of points used to evaluate the maxima
                (at least 10**4)
NPTM      10000
C* --- pi0pi0 (FSKIND=0) or pi+pi- (FSKIND=1) channel
FSKIND  1
C* --- type of radiation
= (1 1))
C*KIND 1 0    ! initial state radiation on
C*KIND 0 1    ! final   state radiation on
C*KIND 1 1    ! initial+final state radiation
KIND   1 1
C*  --- Model for the Pion form factor
C*FPKIND 1  ! KS
C*FPKIND 2  ! GS
C*FPKIND 3  ! RPT
C*FPKIND 4  ! KS+dual-QCD
FPKIND 3
C*FSR --- brems  f0  phirhopi
rhokind   1       1   0
C*bremkind 1  ! sQED
C*bremkind 2   ! RPT
bremkind   1
C*f0KIND  1  !  linear sigma model
C*f0KIND  2  !  non structure model
C*f0KIND  3  !  chiral unitary approach
C*f0KIND  4  !  kaon loop with f0 only
C*f0KIND  5  !  kaon loop with f0+\sigma
f0KIND 5
C*respikind 1 ! only \rho\pi\gamma
C*respikind 2 ! only \omega\pi\gamma
C*respikind 3 ! both \rho\pi\gamma and \omega\pi\gamma
respikind 3
C* --- Parameters of the Pion Form factor
C*RPT MRHO GAMMARHO MRHOL GRHOL MOMEGA GOMEGA AL BE argA
pi\_rho FV FV1 GV1
FP 0.774 0.143 1.37 0.51 0.7827 8.68E-3 0. 0. 0.
-0.0027 0.154  0.  0.
C*FPIHIGH c_0_pion  c_2_pion  c_3_pion  m_rho2_pion  g_rho2_pion
m_rho3_pion g_rho3_pion
FPIHIGH 1.171 0.011519 -0.0437612 1.7 0.24 2.0517 0.41
C* --- Parameters of Scalar contribution (f0,f0+sigma) in FSR
C*F0+SIGMA Gf0\_k+k- Gf0\_p+p- GPHI phase(deg) f0MASS msigma
gsigpp gsigkk Cf0sig
F0PAR 4.02 -1.76 4.482 0. 0.982 415.0e-3 -2.2  -0.37  0.015
C*f0phase F0+SIGMA m0k m2k LambdaK b0p b1p b2p Lambdap
F0fase   0.363 0.757 1.24 5.4 3.7 5.0 0.200
C*F0nonstr  gf0_pp gf0_kk gphi_f0g gphi_sgg a0
F0nonstr  0.57 1.14  3.5  5.0e-3 1.
C*-------- Parameters of double resonance mechanism
C*VMDPAR g\_rhopig g\_phirpi prhores beta\_bro beta\_wphi
c\_modrpi c\_pharpi c\_modwpi c\_phawpi
VMDPAR 0.295d0 0.811d0 0.677d0 32.996d0 163.0d0
0.26d0 3.1112d0 0.85019d0 0.45d0

C* --- histo output file
*HOUT    'isr_rpt_fsr_sqed_f0+sig_50_130_1e6.hbk'
ENES 1.039202865
GMIN   0.02
GMAX   0.600
C*ACUT theta_min_\gamma theta_max_\gamma theta_min_\pi theta_max_\pi
ACUT   50. 130. 50. 130
C* ---------------------------------------
STOP
END
\end{verbatim}
\end{appendix}

\begin{appendix}
\section*{Appendix C. Output file}\label{App_B}
\setcounter{equation}{0}
\def\theequation{C.\arabic{equation}}

There are two kinds of output files:
(\textbf{output.hbk}, HBOOK/PAW ntuple format, defined in \textbf{cards\_fasterd.dat}) contains
the 4-momenta of outgoing particles for each accepted event;
 (\textbf{output.out}, text format, output of the program) contains
the number of generated and accepted events, the value of the
cross section with the error, and the biggest value of the integrand.

\end{appendix}


\begin{thebibliography}{99}

\bibitem{Upgm} G.W.~Bennett {\it et al.}, Phys. Rev. D {\bf 73} (2006) 072003;  \\
James P. Miller, Eduardo de Rafael, B. Lee Roberts, Rep. Prog. Phys. 70 (2007) 795 .

\bibitem{amu_th_exp}
M.~Davier, S.~Eidelman, A.~H\"{o}cker and Z.~Zhang,  Eur. Phys. J. C \textbf{31}(2003) 503; \\
K.~Hagiwara, A. D.~Martin, D.~Nomura, T.~Teubner,  Phys. Rev. D \textbf{69} (2004) 093003; \\
F.~Jegerlehner, Acta Phys. Polon. B \textbf{38} (2007) 3021.


\bibitem{eid_jeg}
 S.~Eidelman and F.~Jegerlehner,
  %``Hadronic contributions to g-2 of the leptons and to the effective fine
  %structure constant alpha (M(z)**2),''
  Z.\ Phys.\ C {\bf 67} (1995) 585.

\bibitem{kluge} W.~Kluge, arXiv:hep-ex/0805.4708.


\bibitem{Chen_75}  Min-Shin Chen and P.M. Zerwas, Phys. Rev. D \textbf{11} (1975) 58.

\bibitem{Rr1}  A.B. Arbuzov, E.A. Kuraev, N.P. Merenkov and L. Trentadue, JHEP
\textbf{12} (1998) 009;  \\ M. Konchatnij and N.P. Merenkov, JETP
Lett. \textbf{69} (1999) 811.

\bibitem{Rr2}  S. Binner, J.H. K\"uhn, K. Melnikov, Phys. Lett. B \textbf{%
459} (1999) 279;%\\ S. Spagnolo, Eur. Phys. J. C \textbf{6}
(1999) 637; \\J. K\"uhn, Nucl. Phys. B (Proc. Suppl.) \textbf{98} (2001) 2.


\bibitem{Baier_65} V.N. Baier and V.A. Khoze, Sov. Phys. JETP \textbf{21}
(1965) 629; \textit{ibid.}, \textbf{21} (1965) 1145.


\bibitem{Khoze_02}  V.A.~Khoze, M.I.~Konchatnij, N.P.~Merenkov {\it et al.} Eur. Phys. J. C \textbf{25} (2002) 199.                        .

\bibitem{kloe} A.~Aloisio {\it et al.}  [KLOE Collaboration],
  %``Measurement of sigma(e+ e- $\to$ pi+ pi- gamma) and extraction of  sigma(e+%e- $\to$ pi+ pi-) below 1-GeV with the KLOE detector,''
  Phys.\ Lett.\ B {\bf 606} (2005) 12.

\bibitem{kloe_large} D. Leone, Nucl. Phys. B - Proc. Suppl. {\bf 162} (2006) 95.

\bibitem{Pancheri:2007xt}
  G.~Pancheri, O.~Shekhovtsova and G.~Venanzoni,
  %``Final state radiation and a possibility to test a pion-photon   interaction
  %model near two-pion threshold,''
  J.\ Exp.\ Theor.\ Phys.\  {\bf 106} (2008) 470; \\
    G.~Pancheri, O.~Shekhovtsova and G.~Venanzoni,
  %``Test of FSR in the process e+ e- --> pi+ pi- gamma at DAFNE and extraction
  %of the pion form factor at threshold,''
  Phys.\ Lett.\  B {\bf 642} (2006) 342.
    %%CITATION = PHLTA,B642,342;%%




\bibitem{czyz} H.~Czy\.z, A.~Grzelinska and J.~H.~K\"uhn,
  %``Charge asymmetry and radiative Phi decays,''
  Phys.\ Lett.\ B {\bf 611} (2005) 116.

\bibitem{phokhara} G.~Rodrigo
H.~Czy\.z, J.~H.~K\"uhn and M.~Szopa,
    Eur.\ Phys.\ J.\ C {\bf 24} (2002) 71; \\
H.~Czy\.z, A.~Grzelinska, J.~H.~K\"uhn and G.~Rodrigo,
  %``The radiative return at Phi- and B-factories: Small-angle photon emission
  %at next to leading order,''
  Eur.\ Phys.\ J.\ C {\bf 27} (2003) 563; \\
H.~Czy\.z, A.~Grzelinska, J.~H.~K\"uhn and G.~Rodrigo,
  %``The radiative return at Phi- and B-factories: Small-angle photon emission
  %at next to leading order,''
  Eur.\ Phys.\ J.\ C {\bf 33} (2004) 333.


\bibitem{graz}
  K.~Melnikov, F.~Nguyen, B.~Valeriani and G.~Venanzoni,
  %``Contribution of the direct decay phi $\to$ pi+ pi- gamma to the process  e+  %e- $\to$ pi+ pi- gamma at DAPHNE,''
  Phys.\ Lett.\ B {\bf 477} (2000) 114.



\bibitem{kuhn_ff} J.~H.~K\"uhn and A.~Santamaria,
  Z.\ Phys.\ C {\bf 48} (1990) 445.

\bibitem{gounaris_ff} G.~J.~Gounaris, J.~J.~Sakurai, Phys. Rev. Lett. \textbf{21} (1968) 244.

\bibitem{kuhn_impr} Ch.~Bruch, A.~Khodjamirian, J.~H.~K\"uhn, Eur.
Phys. J. C {\bf 39} (2005) 41.


\bibitem{Ecker_89} G. Ecker, J. Gasser, A. Pich and E. de Rafael, Nucl.
Phys. B \textbf{321} (1989) 311; \\ G. Ecker, J. Gasser, H.
Leutwyler, A. Pich and E. de Rafael, Phys. Lett. B \textbf{223} (1989) 425.


\bibitem{domin} C.~A.~Dominguez, Phys. Lett. B {\bf 512} (2001) 331.


\bibitem{our_2005}  S.~Dubinsky, A.~Korchin, N.~Merenkov, G.~Pancheri and O.~Shekhovtsova,
  %``Final-state radiation in electron positron annihilation into a pion
  %pair,''
  Eur.\ Phys.\ J.\ C {\bf 40} (2005) 41.



\bibitem{gino}  G.~Isidori, L.~Maiani, M.~Nicolaci and S.~Pacetti,
  %``The e+ e- $\to$ P(1) P(2) gamma processes close to the phi peak: Toward a
  %model-independent analysis,''
   JHEP {\bf 0605} (2006) 049 .

\bibitem{bramon} A.~Bramon, G.~Colangelo and M.~Greco, Phys. Lett. B {\bf 287} (1992) 263.

\bibitem{lucio} J.~L.~Lucio, M.~Napsuciale, Phys. Lett. B {\bf 331} (1994) 418.

\bibitem{chpt_phi} E.~Marco, S.~Hirenzaki, E.~Oset, H.~Toki, Phys. Lett. B {\bf 470} (1990) 20; \\
J.~A.~Oller, Phys. Lett. B {\bf 426} (1999) 20.


\bibitem{ach_sol} N.~N.~Achasov, V.~V.~Gubin and E.~P.~Solodov,
  %``Interference in the reaction e+ e- $\to$ gamma pi+ pi- and searches  for
  %the decay Phi $\to$ gamma f(0) $\to$ gamma pi+ pi-,''
  Phys.\ Rev.\ D {\bf 55} (1997) 2672;\\
  N.~N.~Achasov and V.~V.~Gubin,
  %``Interference in the reaction e+ e- $\to$ gamma pi+ pi- and the final  state  %interaction,''
  Phys.\ Rev.\ D {\bf 57} (1998) 1987.


\bibitem{achasov}N.N.~Achasov and A.V.~Kiselev,
  Phys.\ Rev.\ D {\bf 73} (2006) 054029.


\bibitem{close}
  F.~E.~Close, N.~Isgur and S.~Kumano,
  %``Scalar mesons in phi radiative decay: Their implications for spectroscopy
  %and for studies of CP violation at phi factories,''
  Nucl.\ Phys.\ B {\bf 389} (1993) 513.

\bibitem{kloepi0}
F.~Ambrosino {\it et al.}  [KLOE Collaboration],
  %``Dalitz plot analysis of e+ e- --> pi0 pi0 gamma events at s**(1/2) ~=
  %M(Phi) with the KLOE detector,''
  Eur.\ Phys.\ J.\  C {\bf 49} (2007) 473.

\bibitem{giovannella_pi0} S.~Giovannella, S.~Miscetti, KLOE NOTE
212 (2006), http://lnf.infn.it/kloe/pub/knote/kn212.ps; \\ S.~Giovannella, S.~Miscetti, KLOE NOTE 213 (2006), http://lnf.infn.it/kloe/pub/knote/kn213.ps.

\bibitem{cern_lib} CERN Program Library, Library MATHLIB, V113; \\
F.~James, Computer Phys. Comm. {\bf 60} (1990) 329.

\end{thebibliography}
\end{document}